\newcommand{\ii}{\'{\i}}
\documentclass{ws-mpla}

\begin{document}

\markboth{E. Hern\'andez, J. Nieves, M. Valverde}
{Neutrino induced weak pion production off the nucleon}

%%%%%%%%%%%%%%%%%%%%% Publisher's Area please ignore %%%%%%%%%%%%%%
\catchline{}{}{}{}{}
%%%%%%%%%%%%%%%%%%%%%%%%%%%%%%%%%%%%%%%%%%%%%%%%%%%%%%%%%%%%%%%%%%%

\title{\vspace{-1cm}NEUTRINO INDUCED WEAK PION PRODUCTION OFF THE NUCLEON
%\TeX\ OR \LaTeX\footnote{For the title, try not to use more than 
%three lines. Typeset the title in 10 pt Times Roman, uppercase and 
%boldface.}
}

\author{\footnotesize \vspace{-.5cm}E. HERN\'ANDEZ%\footnote{
%Typeset names in 8 pt Times Roman, uppercase. Use the footnote to 
%indicate the present or permanent address of the author.}
}

\address{Departamento de F\ii sica Fundamental e IUFFyM, Universidad de Salamanca\\%, Plaza de la Merced s/n\\
37008 Salamanca, Spain\\
%Country\footnote{State completely without abbreviations, the 
%affiliation and mailing address, including country and e-mail address. 
%Typeset in 8 pt Times Italic.}\\
gajatee@usal.es}

\author{J. NIEVES and M. VALVERDE}

\address{Departamento de F\ii sica At\'omica, Molecular y Nuclear, Universidad de Granada\\%, Avda. Fuente Nueva s/n\\
18071 Granada, Spain
}
%\author{M. VALVERDE}
%
%\address{Departamento de F\ii sica At\'omica, Molecular y Nuclear, Universidad de Granada\\%, Avda. Fuente Nueva s/n\\
%18071 Granada, Spain
%}
\maketitle

\pub{Received (Day Month Year)}{Revised (Day Month Year)}

\begin{abstract}
We study neutrino induced one-pion production  off the nucleon in and around the Delta resonance region.
Apart from  the Delta-pole mechanism we  include  background terms required by chiral symmetry. These
background terms give sizeable contributions in all channels. %As a result, and in order 
To better reproduce the 
ANL $q^2$-differential cross section  data, we  make a new fit of the   $C_5^A(q^2)$ axial nucleon to Delta form factor. 
The new result $C_5^A(0)=0.867\pm 0.075$ is some 30\% smaller  than the commonly accepted value.
%the off-diagonal Goldberger-Treiman relation prediction.
 This correction is compatible with most quark model estimates and a
recent lattice calculation~\cite{alexandrou07}.% by C. Alexandrou {\it et al.}~\cite{alexandrou07}.  
%For the axial mass we get $M_{A\Delta}=0.985\pm0.082$\,GeV
%which represents only a few percent reduction over the commonly accepted value.

\end{abstract}

\ccode{PACS Nos.: 25.30.Pt, 13.15.+g, 12.15.-y, 12.39.Fe.}
%
%
%
%
%\vspace*{-.25cm}
\section{Introduction and brief description of the model}	
\begin{figure}[h!!!]
\centerline{\psfig{file=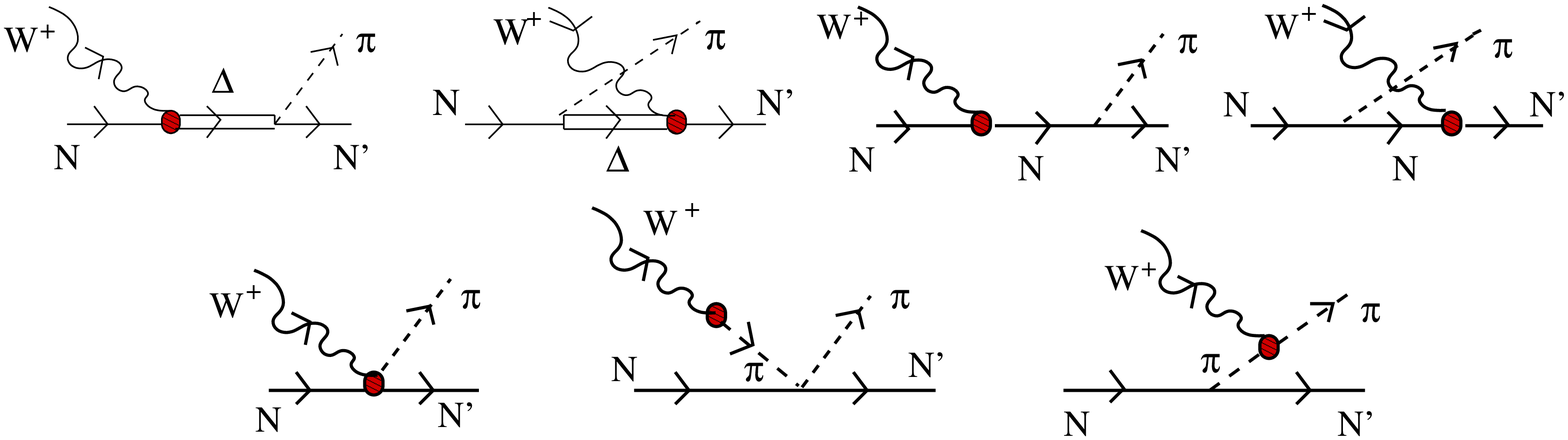,width=4.in}}
\caption{\footnotesize Model for the $W^+N\to N^\prime\pi$
  reaction. We have direct and crossed
  $\Delta(1232)$  and nucleon  pole terms,
  contact and pion pole contribution,  and finally the
  pion-in-flight term.  }
  \label{fig:diagramas}
\end{figure}

Pion production reactions induced by neutrinos are very interesting as a means to study hadronic 
structure. Besides, they have become very relevant in the analysis of neutrino oscillation experiments where pion production 
contributes to the background. 
%While neutrino oscillation experiments generally use complex nuclei as targets and are thus  affected by medium
%effects,  a good understanding of the fundamental reaction at the nucleon level is needed. This is the purpose of this
%work. 
Most previous studies\cite{adler,lle,previous_noback,paschos,lala} of pion production processes at intermediate energies considered 
only the Delta pole ($\Delta P$)
mechanism (See Fig.\ref{fig:diagramas}).
% in which the neutrino excites a $\Delta(1232)$ resonance that subsequently decays into $N\pi$. 
Here, we shall also include
background terms required by chiral symmetry. Some background terms have been included before in the works of Ref.\cite{previous_back}
although they are not fully consistent with chiral counting rules.
%Due to the lack of space we can not give a full account of our model and all the results we have obtained. Interested readers
%can find all the relevant information in Ref.\cite{hnv07}.
%
%
%
%

For charged current (CC) processes our model includes all contributions depicted in Fig.~\ref{fig:diagramas}. We have
 the $\Delta P$ terms and background terms which
include nucleon-pole terms ($NP$),  contact term ($CT$), pion-pole term ($PP$), and pion in flight term ($PF$). For that purpose we
use a SU(2) nonlinear $\sigma-$model with nucleon and pion degrees of freedom.
%that includes a nucleon doublet and the triplet of pions. 
On top of the vertices and currents
provided by this model we include phenomenological form factors for the weak $NN$ vertex that we take from the work of 
Ref.\cite{galster}. A different form factor is included in the purely axial $PP$ term to account for the $\rho-$meson dominance of
the  $\pi\pi NN$ vertex. Vector current conservation (CVC) and partial conservation of the axial current (PCAC)  require
the  introduction of corresponding form factors in the $PF$ term, and the vector and axial parts of the $CT$ term.
\vspace{-.25cm}
The weak nucleon-Delta vertex is parametrized in terms of four vector and four axial form factors\cite{lle}. We
 use the set of vector form factors of Ref.\cite{lala} determined from the analysis of photo and electroproduction data. Initially we shall 
 take the set of axial form factors in Ref.\cite{paschos}. In this latter work Adler's model\cite{adler} is used to fix $C_3^A=0$ and
 $C_4^A=-C_5^A/4$, while $C_6^A$ is obtained from $C_5^A$ by PCAC. For $C_5^A$ they use the parameterization\vspace{-.15cm}
 \begin{equation}
 C_5^A(q^2)=\frac{C_5^A(0)}{(1-\frac{q^2}{M^2_{A\Delta}})(1-\frac{q^2}{3M^2_{A\Delta}})}\ \ ; \ \ {\rm with}\     C_5^A(0)=1.2,\ \
 M_{A\Delta}=1.05\,{\rm GeV} 
 \label{eq:c5a}              
 \end{equation}
 $C_5^A(0)=1.2$ corresponds to the value obtained from the $g_{\pi N \Delta}$
 coupling constant using  
 the off-diagonal Goldberger-Treiman relation.
 We shall call this Set I. % and as we explain in what follows this will not be  our preferred choice.

Due to the lack of space we can not give a full account of our model and all the results we have obtained. Interested readers
can find all the relevant information in Ref.\cite{hnv07}.
%\vspace*{-.475cm}
\section{Results and discussion}
\begin{figure}[h!!!]
%\centerline{\includegraphics[height=6cm]{anl.epsi}}
\centerline{\psfig{file=anl.epsi,width=2.45in}\psfig{file=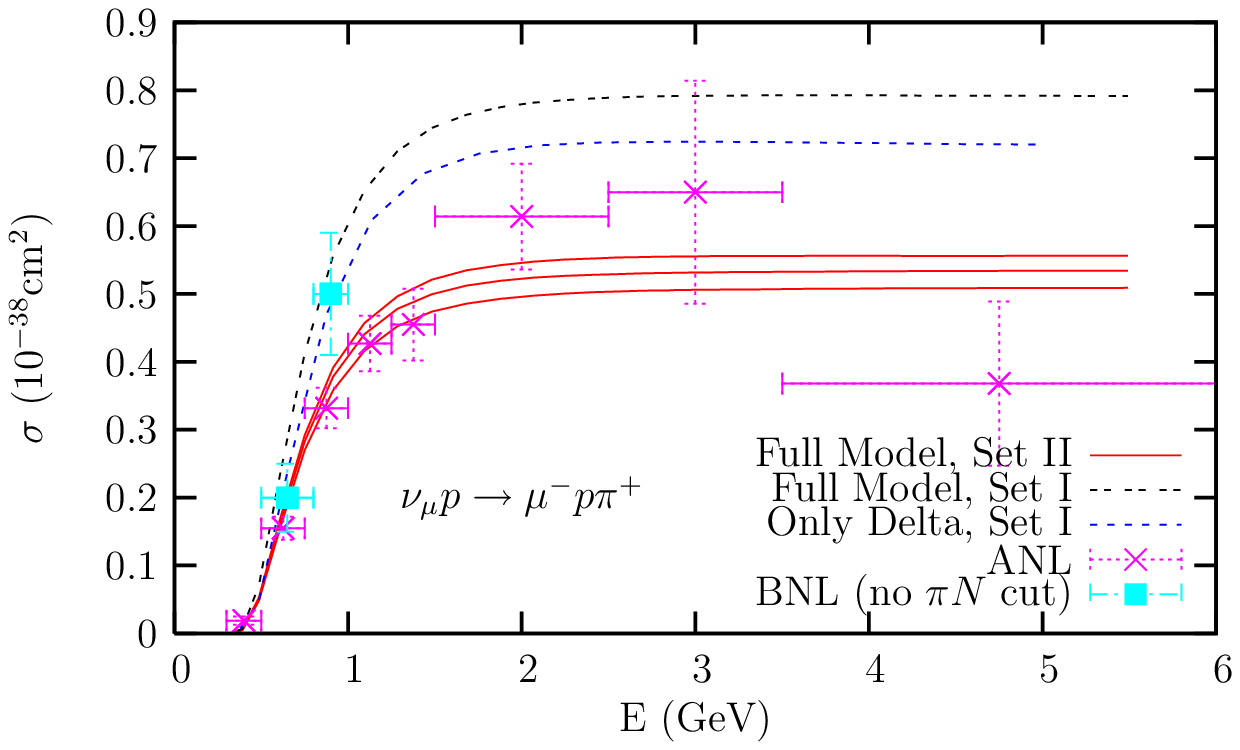,width=2.45in}}
\centerline{\psfig{file=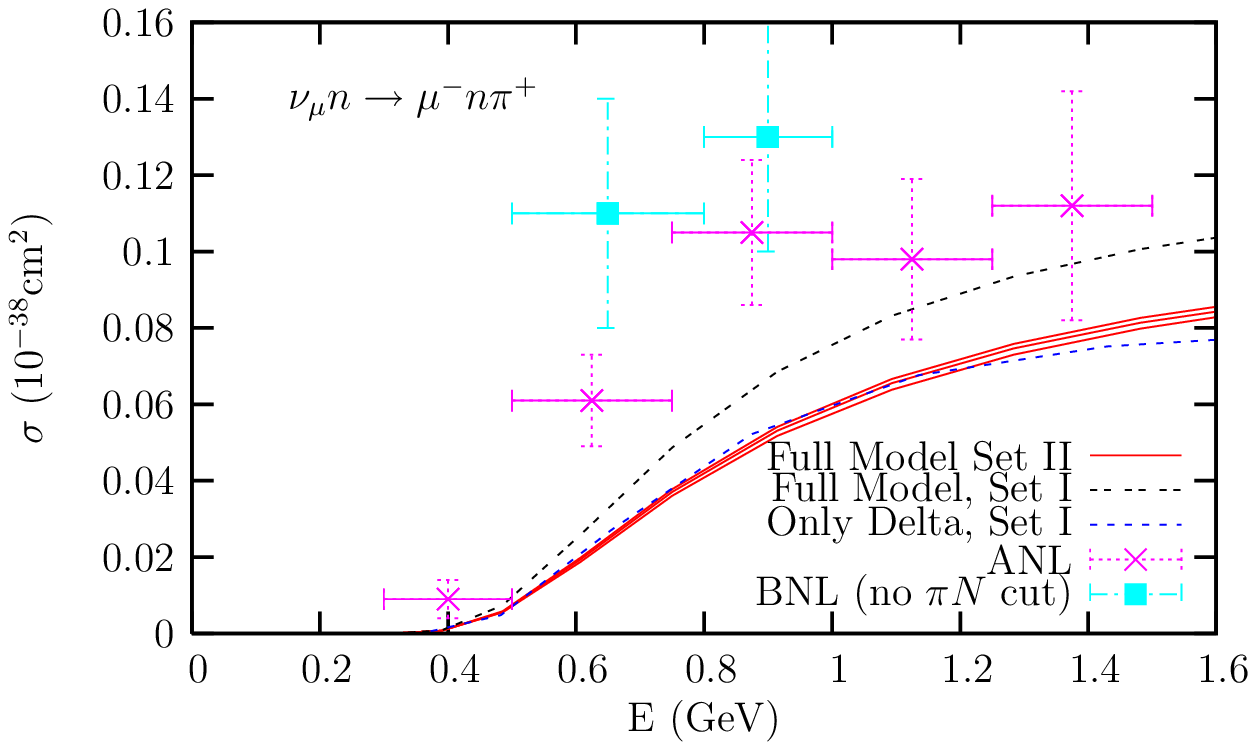,width=2.45in}\psfig{file=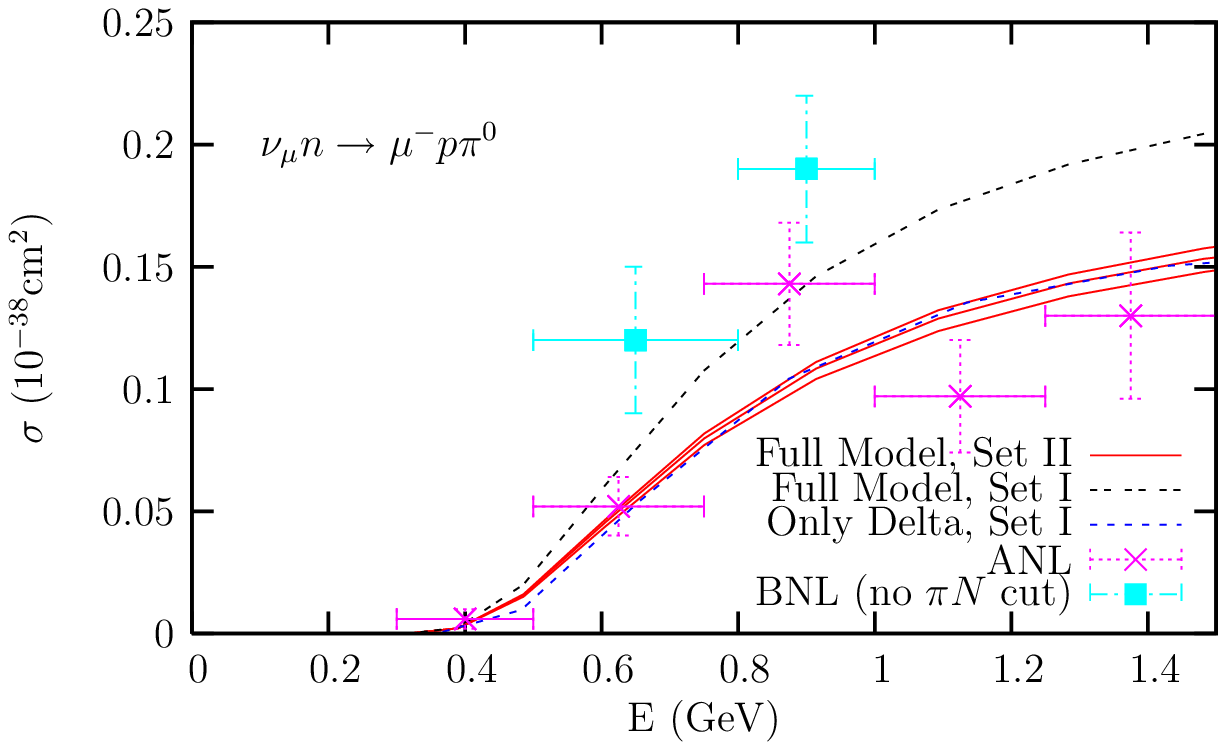,width=2.45in}}
\centerline{\psfig{file=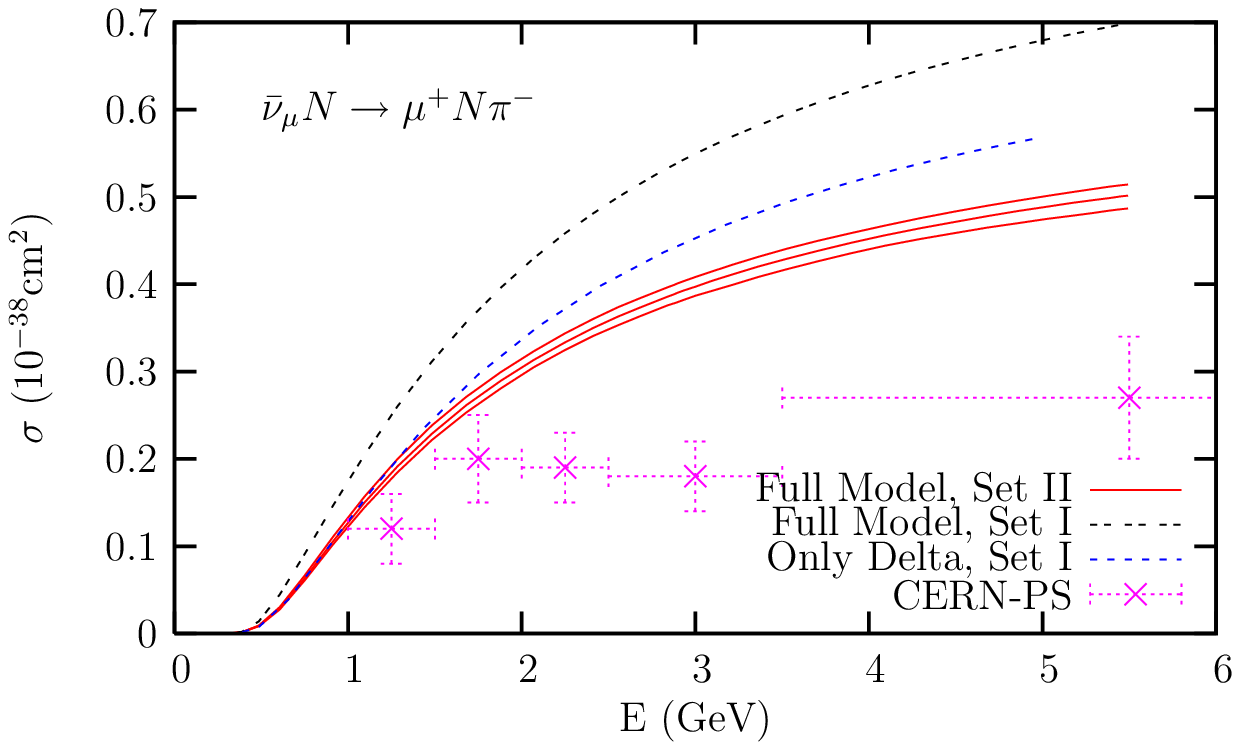,width=2.45in}\psfig{file=derrick.epsi,width=2.45in}}
\centerline{\psfig{file=s_ppi0.epsi,width=2.45in}\psfig{file=anti.epsi,width=2.45in}}%\centerline{\psfig{file=erange5.epsi,width=3.0in}}
\vspace*{8pt}
\caption{%\footnotesize 
Top left panel: Flux-averaged $q^2$-differential  cross section
$\int_{M+m_\pi}^{1.4{\rm GeV}}dW\frac{d\bar\sigma_{\nu_\mu \mu^-}}{dq^2dW}$ for \hbox{$\nu_\mu p\to \mu^- p\pi^+$}. $W$ stands for 
the final pion-nucleon invariant mass.
Other panels : different charged current and neutral current reactions results. We compare our results with 
ANL and BNL  data. With the exception of the $\nu n\to \nu p\pi^-$ reaction, ANL data has a cut at $W=1.4$\,GeV which we have 
also implemented in our calculation.
 }
\label{fig:neutrino}
\end{figure}

In the top left panel of Fig.\ref{fig:neutrino} we show the flux-averaged $q^2$-differential $\nu_\mu p\to \mu^- p\pi^+$ cross section
$\int_{M+m_\pi}^{1.4{\rm GeV}}dW\frac{d\bar\sigma_{\nu_\mu \mu^-}}{dq^2dW}$, where $W$ is the invariant mass of the final hadronic state.
The inclusion of the background terms spoils the agreement with ANL data if Set I is used. The least well known ingredients of the
model are the axial nucleon-Delta form factors, of which $C_5^A$ gives the largest contribution. This strongly suggests a new fit of 
 $C_5^A$  to the ANL experimental data. Keeping the parameterization  in Eq.(\ref{eq:c5a}) we obtain the new values
\begin{equation}
  C_5^A(0)=0.867\pm0.075,\ \ M_{A\Delta}=0.985\pm0.082\,{\rm GeV} 
 \label{eq:c5a0}              
 \end{equation}
that we shall call Set II. The new $C_5^A(0)$ value is some 30\% smaller than the commonly assumed one, but it is not in contradiction
with a recent lattice determination\cite{alexandrou07}.
In the panel  we also show  a 68\% confidence level band deduced from the statistical errors quoted above.
In the rest of the panels we show the results for different CC and neutral current (NC) driven processes. In all of them the
background terms give sizeable contributions. The use of Set II of axial nucleon-Delta form factor increase the overall agreement
with data. 
For the neutrino induced CC processes we also show  BNL data\cite{bnl} which seem to  favor a larger $C_5^A(0)$ value.
For antineutrino induced CC processes we see our full model calculation with Set II  gives the best results but it is still
larger than the experimental data obtained at CERN\cite{cern}. A recent calculation\cite{athar} claims that medium and pion absorption effects
can perfectly explain this discrepancy between  theoretical results on the nucleon and  antineutrino experimental data actually measured in a 
freon-propane target. 

%\begin{figure}[tbh]
%\centerline{\psfig{file=derrick.epsi,width=2.25in}}
%%\centerline{
%\psfig{file=s_ppi0.epsi,width=2.25in}\psfig{file=anti.epsi,width=2.25in}
%\centerline{\psfig{file=erange5.epsi,width=3.0in}}
%\vspace*{8pt}
%\caption{\footnotesize Neutral current reactions results. }\label{fig:diagramas}
%\end{figure}
The last three panels of Fig~\ref{fig:neutrino}  show results for NC
processes. The isovector contributions to these can be obtained directly from the
CC ones using isospin symmetry. Besides, there are two different isoscalar contributions, one given in terms of the isoscalar part of the
electromagnetic current and another one related to the $s\bar s$ content of the nucleon. Both are of the $NP$ type and their
expressions and the form factors used in this case are given in Ref.\cite{hnv07}.  Our full model using Set II reproduces
 fairly well the experimental results for $\nu n\to \nu p\pi^-$. The bottom left panel shows our results
results are not sensitive to  the $s\bar s$ content of the nucleon. Last panel serves two
purposes. First, it exemplifies the fact that antineutrino induced NC processes are suppressed with respect to neutrino induced ones, and
second, it shows there is little isovector-isoscalar interference. In fact the isovector contribution is dominant.
%\centerline{\psfig{f

\vspace*{-.22cm}
\section*{Acknowledgments}
 This research was supported by DGI and FEDER funds, under contracts
FIS2005-00810,  %FPA2004-05616, 
FIS2006-03438 and FPA2007-65748, by J. %unta de
de Andaluc\'\i a and J. %unta 
de Castilla y Le\'on under contracts FQM0225
and SA016A07, and it is part of the EU integrated infrastructure
initiative Hadron Physics Project under contract number
RII3-CT-2004-506078.

\end{document}